\documentclass[traditabstract]{aachanged}
\usepackage[english]{babel}
\usepackage{longtable}
\usepackage{graphicx}	
\usepackage{epsfig}
\usepackage{epsf}
\usepackage{amsmath}	
\usepackage{amssymb}	
\usepackage{xcolor,soul}
\usepackage{subcaption}
\usepackage{soul}
\usepackage[normalem]{ulem}

\begin{document}

\title{\bf Determination of HII region metallicity in the context of estimating the primordial helium abundance.}
\author{O.A. Kurichin$^{1}$\thanks{E-mail: o.chinkuir@gmail.com}, P.A. Kislitsyn$^1$, A.V. Ivanchik$^1$}
\authorrunning{Kurichin et al.}
\date{Accepted September 07, 2021}

\institute{\it{$^{1}$Ioffe Institute, Russian Academy of Sciences, ul. Politekhnicheskaya 26,
St. Petersburg, 194021 Russia}\\}

\abstract{The primordial $^4$He abundance (Y$_p$) is one of the key characteristics of Primordial Nucleosynthesis processes that occurred in the first minutes after the Big Bang. Its value depends on the baryon/photon ratio $\eta \equiv n_b /n_{\gamma}$, and is also sensitive to the relativistic degrees of freedom which affect the expansion rate of the Universe at the radiation-dominated era. At the moment, the most used method of the determination of Y$_p$ is the study of the metal deficient HII regions located in blue compact dwarf galaxies (BCDs). In this paper, we discuss in detail various methods of the determination of HII region metallicity in the context of Y$_p$ analyses. We show that some procedures used in the methods lead to biases in the metallicity estimates and underestimation of their uncertainties. We propose a modified method for the metallicity determination, as well as an additional criterion for selecting objects. We have selected 69 objects (26 objects with high quality spectra from the HeBCD + NIR database and 43 objects from the SDSS catalog), for which we estimate Y and O/H using the proposed method. We have estimated Y$_p = 0.2470 \pm 0.0020$ which is one of the most accurate estimates obtained up to date. Its comparison with the value Y$_p = 0.2470 \pm 0.0002 $ obtained as a result of numerical modelling of Primordial Nucleosynthesis with the value of $\Omega_b$ taken from the analysis of the CMB anisotropy (Planck mission), is an important tool for studying the self-consistency of the Standard cosmological model (a possible discrepancy between these estimates could be an indicator of a new physics).
The application of the proposed method allows one to more correctly estimate Y$_p$ and the slope $d$Y/$d$(O/H). 
Further analysis of the data from the SDSS catalog can significantly increase the statistics of objects for the regression analysis, which in turn can refine the Y$_p$ estimate.}

\keywords{Early Universe, primordial helium, H II regions.}

\maketitle

\section{Introduction}

Modern observational capabilities allow one to study the Universe from the present time ($z = 0$) up to $z \sim 10$, when the first galaxies begin to form. The study of the cosmic microwave background (CMB) is an even earlier observational ``window'' into the Universe. It makes possible to see what the Universe looked like $\sim$ 400 thousands years after the Big Bang ($z \approx 1100$). Primordial (prestellar) cosmological nucleosynthesis -- a process that occurred in the first minutes after the Big Bang -- is the most distant cosmological process ($ z \sim 10^9-10^7 $) available for the investigation. It has observable and verifiable consequences which allow one to probe the Early Universe.
 
Being an ``inverted'' thermonuclear reactor, where the synthesis of elements proceeds during the expansion and cooling of matter, Primordial Nucleosynthesis leads to the formation of the first light nuclei - D, $^3$He, $^4$He, $^7$Li and etc. Their relative abundance depends on the single parameter $\eta \equiv n_b / n_{\gamma}$ -- the baryon/photon ratio. The subsequent chemical evolution of the Universe leads to changes in the primordial isotopic composition of matter, and the production of heavier elements through the stellar nucleosynthesis. Despite this, there are methods for the determination of the primordial abundances of D, $^4$He and $^7$Li. Such determinations allow to estimate the baryon density of the Universe $\Omega_b$ (one of the key cosmological parameters) using the Primordial Nucleosynthesis numerical codes. Comparison of the value $\Omega_b$ obtained for the era of Primordial Nucleosynthesis (the first minutes after the Big Bang) with the value obtained using the analysis of the CMB anisotropy ($\sim$400 thousand years after the Big Bang) is an important tool for checking the self-consistency of the Standard Cosmological Model (a possible discrepancy could indicate new physics).

The classical method of the determination of the primordial $^4$He mass fraction (Y$_p$) is the observation of the metal deficient HII regions located in blue compact dwarf galaxies (BCDs). In such galaxies the star formation rate is decelerated, therefore the galaxies are relatively chemically unevolved and their chemical composition is close to the primordial one. Since the helium mass fraction can only increase over time due to the stellar nucleosynthesis, there is a correlation between the metallicity of HII regions and their helium abundance. Therefore, estimating the current abundance of $^4$He (Y) and the metallicity (Z) of such objects, one can estimate Y$_p$. Using  the oxygen abundance O/H as a metallicity tracer of objects, one can construct a Y$-$O/H diagram. The value of Y$_p$ can be obtained via the extrapolation of the obtained Y$-$O/H dependence to zero metallicity.

\begin{table}
	\centering
	\caption{Recent estimates of Y$_p$ obtained in various independent researches.}
	\vspace{2mm}
	\def\arraystretch{1.1}
	\label{tab:helium_dets}
	\begin{tabular}{lcll} 
		\hline
		& \hspace{-4mm} $Y_p$ & & ~~~~Reference \\
		\hline
		0.2551  & \hspace{-4mm}$\pm$ & \hspace{-4mm} 0.0022  & Izotov et al. (2014) \\
	    0.2449  & \hspace{-4mm}$\pm$ & \hspace{-4mm} 0.0040  & Aver et al. (2015) \\
		0.2446  & \hspace{-4mm}$\pm$ & \hspace{-4mm} 0.0029  & Peimbert et al. (2016) \\
		0.245   & \hspace{-4mm}$\pm$ & \hspace{-4mm} 0.0070   & Fernandez et al. (2018)\\
		0.243   & \hspace{-4mm}$\pm$ & \hspace{-4mm} 0.0050   & Fernandez et al. (2019)\\
		0.2451  & \hspace{-4mm}$\pm$ & \hspace{-4mm} 0.0026  & Valerdi, Peimbert (2019) \\
		0.2436 & \hspace{-4mm}$\pm$ & \hspace{-4mm} 0.0040   & Hsyu et al. (2020) \\
	    0.2453  & \hspace{-4mm}$\pm$ & \hspace{-4mm} 0.0034  & Aver et al. (2021) \\    
	    0.2462  & \hspace{-4mm}$\pm$ & \hspace{-4mm} 0.0022  & Kurichin et al. (2021) \\
	    0.2448  & \hspace{-4mm}$\pm$ & \hspace{-4mm} 0.0033  & Valerdi et al. (2021) \\
		$\mathbf{0.2470}$  & \hspace{-4mm}$\pm$ & \hspace{-4mm} $\mathbf{0.0020}$  & \bf{\bf{This paper}} \\	 
		\hline
        $^* \mathit{0.2470}$ & \hspace{-4mm} $\pm$ & \hspace{-4mm} $\mathit{ 0.0002}$ & Planck collaboration (2020)\\
		\hline
		\multicolumn{4}{l}{ {\scriptsize{$^*$Note that the estimate of Y$_p$ presented in }}} \\
		\multicolumn{4}{l}{ {\scriptsize{Planck Collaboration (2020) is not a direct measurement of }}}\\
		\multicolumn{4}{l}{ {\scriptsize{the quantity, but is obtained using the numerical codes of }}}\\
		\multicolumn{4}{l}{ {\scriptsize{Primordial Nucleosynthesis with the input parameter $\Omega_b$  }}}\\
		\multicolumn{4}{l}{ {\scriptsize{estimated from the analysis of the CMB anisotropy. }}}\\
	\end{tabular}
\end{table}

Table \ref{tab:helium_dets} presents the recent estimates of the primordial helium abundance obtained using various methods of the analysis of HII regions. Additionally, there are also estimates of Y$_p$ obtained using other approaches. Cooke and Fumagali (2018) analyzed the absorption spectrum of an intergalactic gas (at $z = 1.724$) on the line of sight toward the quasar HS\, 1700+6416. The authors provided the estimate of Y$_p = 0.250 \pm 0.033$. A similar approach is currently widely used to estimate the primordial deuterium abundance (Noterdaeme et al., 2012; Balashev et al., 2016; Riemer-Sørensen et al., 2017; Cooke et al., 2018; Zavarygin et al., 2018). Another way to estimate Y$_p$ is the study of the emission radio lines of helium and hydrogen corresponding to the Rydberg state transitions in the nearby HII regions. Using this approach Tsivilev et al. (2019) put a lower limit Y$_p \ge 0.2519 \pm 0.0115$. A significant advantage of these approaches is the almost complete absence of any systematic effects which affect the estimates of $n($He$)/n($H). On the other hand, these methods still have lower accuracy compared to the classical method for the determination of Y$_p$.

In this paper, we discuss in detail some of the difficulties that arise in the determination of the primordial helium abundance, and propose a way to eliminate them.

\section{Standard method for estimating O/H.}

\noindent
The metallicity of HII region can be determined as the sum of the abundances of all chemical elements heavier than helium. However, the mass fractions of many metals (relative to hydrogen) in such objects are $\lesssim 10 ^ {- 7}$, and the emission lines of these elements are difficult to detect. Therefore, oxygen (one of the most abundant metals in such objects) is usually used as a tracer of  metallicity when determining Y$_p$. The relative abundance of oxygen O/H is defined as the sum of oxygen ionization states (for typical physical conditions in HII regions, such states are OII and OIII, i.e., O/H = OII/H + OIII/H). Abundances of ionization states can be estimated using a two-zone temperature model of HII regions. According to this approach, the forbidden OIII lines originate from the vicinity of an ionizing radiation source which is characterized by the temperature of the interstellar medium electrons $T_e$(OIII). The forbidden OII lines originate from the distant layers of the HII region, characterized by the electron temperature $T_e$(OII). These temperatures can be estimated, using the so-called direct method, from the ratio of forbidden line fluxes for the corresponding ions: $T_e$(OIII) is determined by the ratio of fluxes $ \lambda4363 / (\lambda4959 + \lambda 5007) $, and $T_e$(OII) is determined by the ratio $ (\lambda 7320 + \lambda 7330) / (\lambda 3726 + \lambda 3729) $ (see, i. e., Pilyugin et al. (2010)). Note that weak [OII] lines $\lambda 7320, 7330$ are often not available for observations. Therefore, in papers devoted to the determination of physical conditions in HII regions, $T_e$(OII) is usually estimated indirectly via the relations $T_e$(OII) = $f(T_e$(OIII)). Namely, Izotov et al. (2014), Fernandez et al. (2018), Hsyu et al. (2020), Kurichin et al. (2021), Aver et al. (2021) used various empirical relations $T_e$(OII) = $f(T_e$(OIII)). Only in the papers Valerdi et al. (2019, 2021) for eight HII regions, in spectra of which all of the necessary oxygen lines have been observed, $T_e$(OII) was estimated directly. We also note that in the papers Peimbert et al. (2016) and Valerdi et al. (2019) the authors have carried out a detailed modeling of each studied HII region in order to assess the their physical conditions and metallicity. Such an approach allows to obtain the most accurate estimates of the interstellar medium properties. On the other hand, this approach is strongly dependent on the quality of the studied spectrum. Therefore it is unsuitable for the analysis of large spectroscopic databases (note that Peimbert et al. (2016) analysed only five objects, and Valerdi et al. (2019) - only one).

In this paper, we show that the use of the relations $T_e$(OII) = $f(T_e$(OIII)) leads to a systematic shift in the estimates of OII/H abundance, which directly affects the determination of  Y$_p$. For the correct determination of OII/H, one needs to assess $T_e($OII) via the direct method using the ratio of [OII] lines $(\lambda3726 + \lambda3727)/(\lambda7320 + \lambda7330)$. Thus, for the determination of  Y$_p$ one should select only objects in spectra of which the required [OII] lines are detected. We see this as a new important selection criterion.

\begin{figure}
\begin{center}
        \includegraphics[scale=0.2]{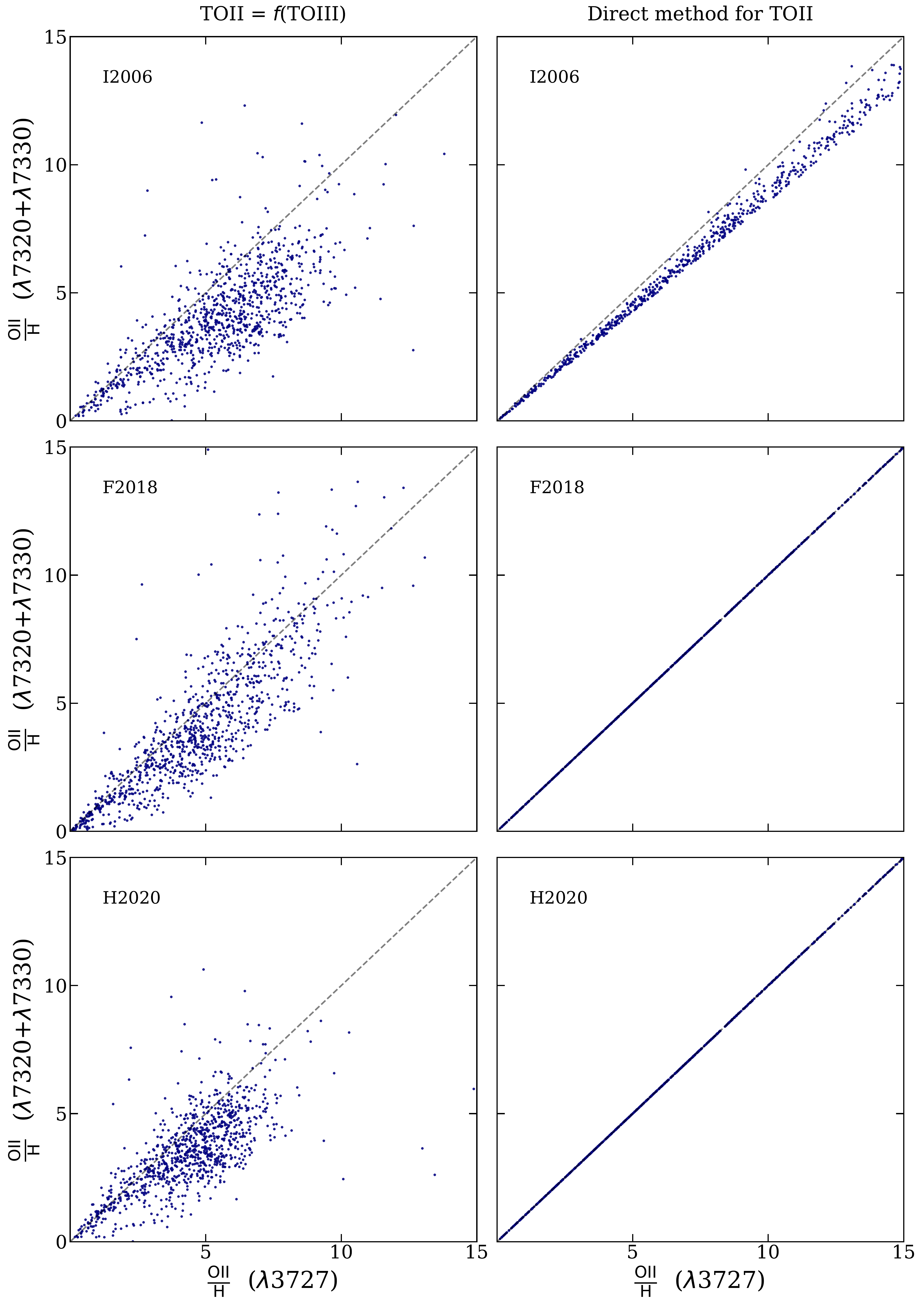}
        \caption{{\small{The distribution of OII/H estimates obtained using the [OII] $\lambda 3727$ and $ \lambda 7320,7330$ lines for objects from the HeBCD + NIR and SDSS databases. The OII/H is estimated using various methods from the literature. The upper panels show the estimates obtained by the method presented in  Izotov et al. (2006). The central panels show the estimates obtained by the method presented in Hsyu et al. (2020). The lower panels show the estimates obtained by the method presented in Fernandez et al. (2018). The figure shows a significant discrepancy in the OII/H estimates if they are obtained using indirect  method for estimating $T_e$(OII) via the formulas $T_e$(OII) = $f(T_e($OIII)) (left panels). The discrepancy is eliminated by using directly estimated $T_e$(OII) (right panels).}}}
        \label{Oh+_descrepancy}
\end{center}
\end{figure}

\subsection*{Issues with the standard method}

The detailed study of the determination of OII/H shows that the OII/H value estimated using the [OII] $\lambda3727$ line is inconsistent with the same value estimated using the [OII] $\lambda7320, 7330$ lines, and the discrepancy grows with increasing metallicity (see. left panels of fig. \ref{Oh+_descrepancy}). The figure 1 shows OII/H($\lambda 3727$) and OII/H($\lambda 7320, 7330$), estimated for the objects taken from the HeBCD + NIR databases (Izotov et al. (2007, 2014)) and the combined database of the objects from Hsyu et al. (2020) and Kurichin et al. (2021), which were selected from the SDSS catalog. The left panels show the determinations of OII/H obtained using the methods described in the previous paragraph ($T_e$(OII) being estimated indirectly via the relations $T_e$(OII) = $f(T_e$(OIII))). The right panels show the determinations of OII/H obtained using $T_e$(OII) estimated via the direct method. To estimate the temperature by the direct method, we use the PyNeb software package (Luridiana et al., 2015). The package calculates $T_e$(OII) based on modeling the statistical equilibrium of the electron shells at the known ratio of the fluxes of forbidden lines. The comparison of the left and right panels shows that the discrepancy is due to the incorrectly estimated temperature of the low ionization zone. The right panels demonstrate that the direct determination method leads to consistent estimates of metallicity obtained for the different lines. Note that the use of directly estimated $T_e$(OII) for the determination of OII/H does not completely eliminate the discrepancy between the OII/H($\lambda 3727$) and OII/H($\lambda 7320, 7330$) obtained using formulas from Izotov et al. (2006). This probably may be because this method uses formulas for the determination of O/H which are derived based on old atomic data (see Izotov et al. (2006) for details). At the same time, two other methods (central and lower panels) utilize the PyNeb software package (Luridiana et al., 2015) for calculating the metallicity (PyNeb uses up-to-date atomic data). It leads to very good consistency between the OII/H estimates obtained using different OII lines.

\section{Corrected method for estimating  O/H}


\begin{figure}
\begin{center}
        \includegraphics[scale=0.3]{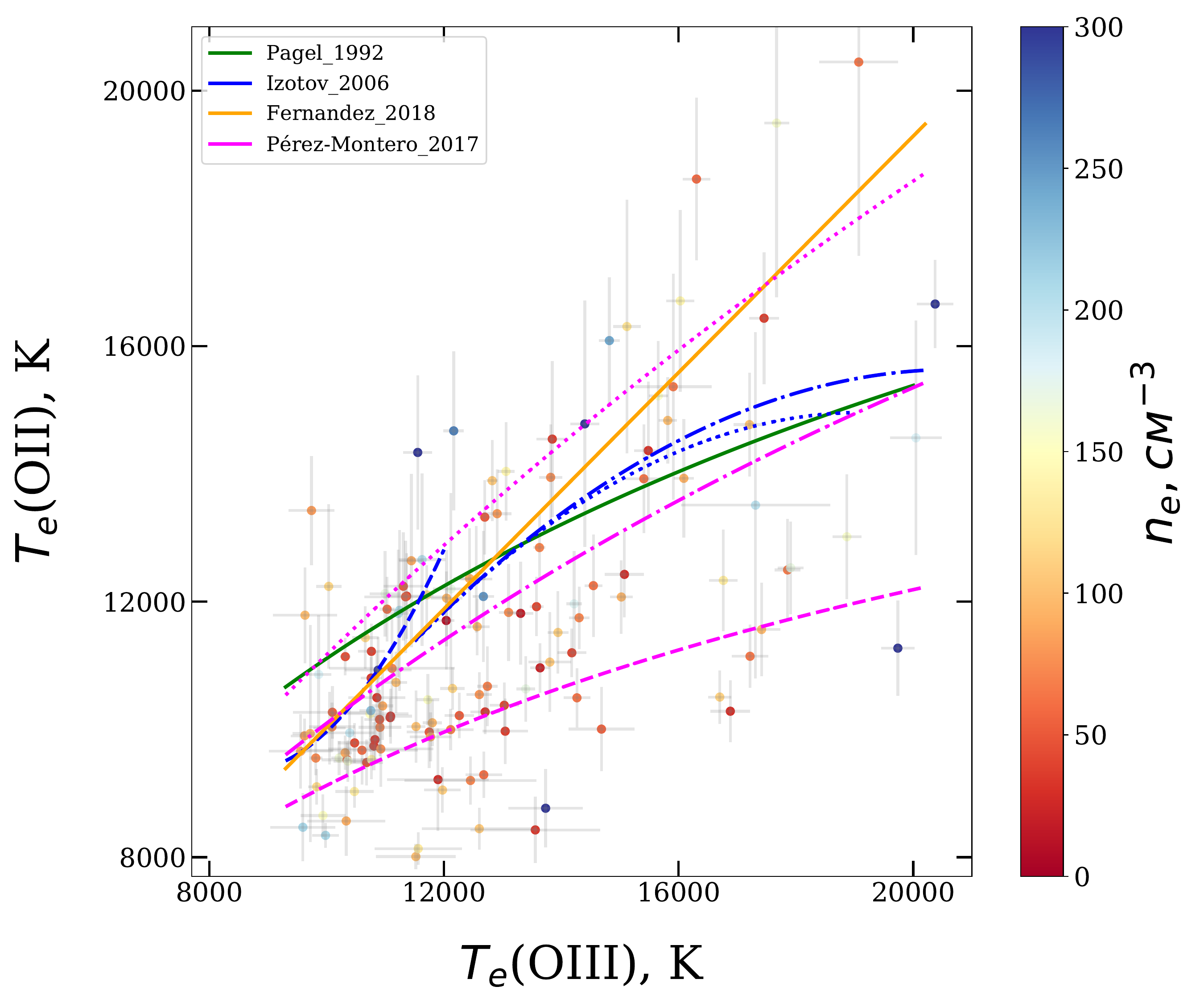}
        \caption{\rm The distribution of $T_e$(OII) and $T_e$(OIII) estimates for HII regions from the SDSS catalog and the HeBCD database. The points are colored according to the estimated $n_e$. Solid lines represent the fitting relations $T_e$(OII) = $f(T_e$(OIII)) from Pagel et al. (1992), Fernandez et al. (2018). The dash-dotted, dotted and dashed blue lines correspond to the relations for low, intermediate, and high metallicity (Izotov et al. (2006)). The dashed, dash-dotted, and dashed magenta lines correspond to the relations from Pérez-Montero (2017) for $n_e = 30, 120, 300~ \mathrm{cm}^{-3}$. 
        }
        \label{to2_to3_ne_7}
\end{center}
\end{figure}

Figure \ref{to2_to3_ne_7} shows the $T_e$(OII) -- $T_e$(OIII) diagram with both temperatures evaluated via the direct method for objects from HeBCD and SDSS databases. In addition, each point is colored corresponding to the estimated electron density in that object. We calculate the temperature and density estimates using the PyNeb software package (Luridiana et al., 2015). The figure also shows various relations $T_e$(OII) = $f(T_e$(OIII)) used in the literature. From figure \ref{to2_to3_ne_7} one can see that there is a certain correlation between $T_e$(OII) and $T_e$(OIII) (the temperature of the low ionization zone increases with an increase in the temperature of the high ionization zone), but none of the relations from literature fully represents it. Similar $T_e$(OII) -- $T_e$(OIII) diagrams for different sets of objects were previously presented in Knyazev et al. (2004), Izotov et al. (2006), Hägele et al. (2006, 2007), and in each case the same strong scatter of points was observed. As a possible solution of the problem, Pérez-Montero (2017) suggested fitting relations which are taking into account the electron density: $T_e$(OII) = $f(T_e($OIII)$,n_e)$. However, as can be seen from figure \ref{to2_to3_ne_7}, these formulas also cannot fully explain the observed scatter (points with high $n_e$ lie both above and below these curves, and similarly for low $n_e$). Therefore, for the determination of the primordial $^4$He abundance, the use of the relations $T_e$(OII) = $f (T_e $(OIII)) is unfavourable. On the other hand, the use of the direct method of estimating $T_e$(OII) gives more reasonable results. 

\begin{figure}
    \centering
    \includegraphics[scale = 0.28]{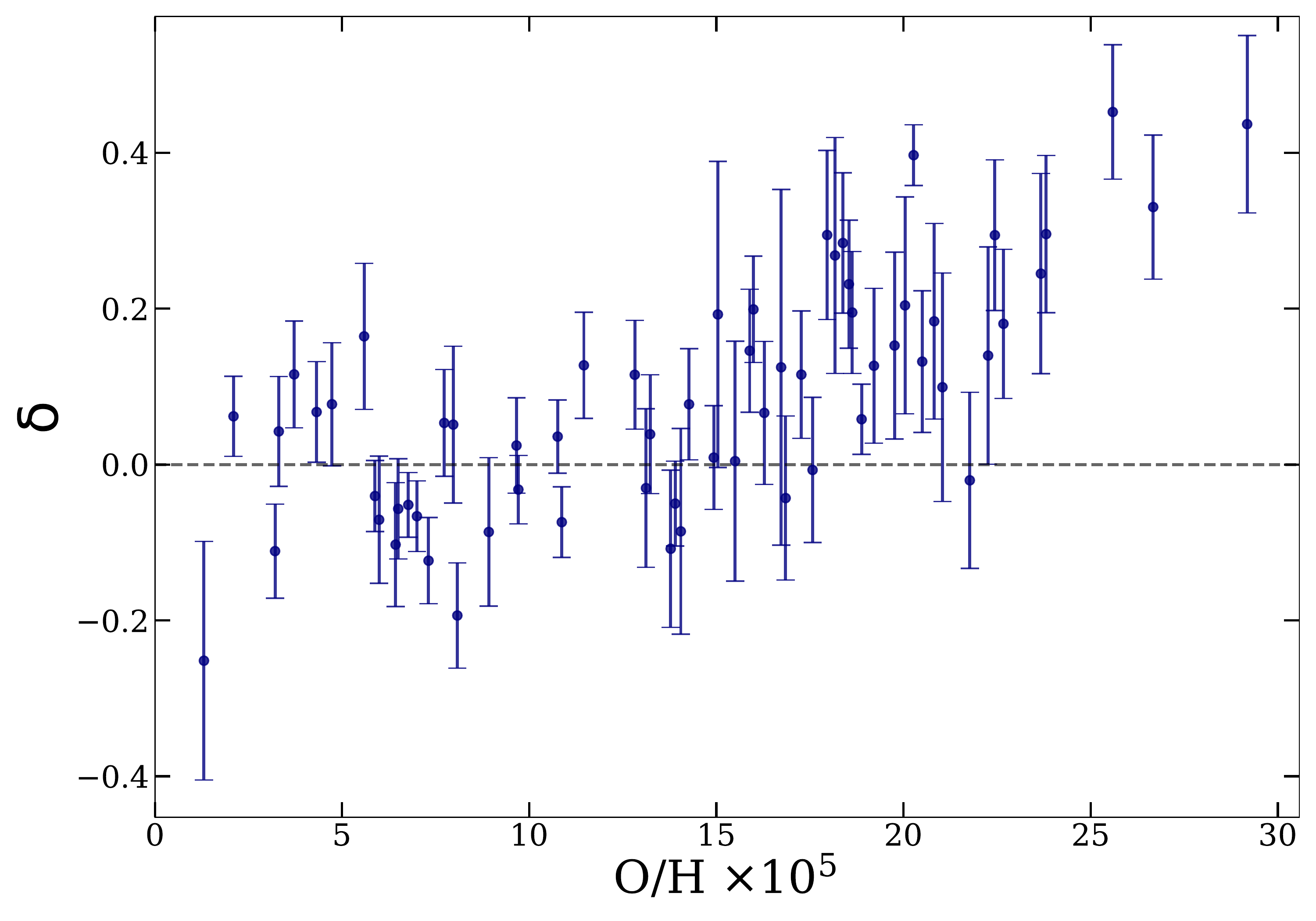}
    \caption{Bias of the estimates of O/H for the objects under study (tables \ref{sdss_obj} and \ref{hebcd_obj}). The value of $\mathrm{\delta}$ is determined via the following relation: $ \mathrm{\delta} = \frac{\rm O/H - O/H_{old}} {\rm O/H} $. Here O/H denotes values obtained using directly estimated $T_e$(OII), and O/H$_ {\rm old}$ denotes values estimated using the relation $T_e$(OII) = $f(T_e$(OIII)). As an example, for the calculation we took the formula from Pagel et al. (1992), which is used in recent paper on the determination of Y$_p$ (Hsyu et al. (2020)).}
    \label{corr_oh}
\end{figure}

The use of any relations between $T_e($OIII) and $T_e($OII) may also lead to underestimation of uncertainties in the estimates of $T_e$(OII). This is because the uncertainty in $T_e$(OII) in this case is entirely determined by the uncertainty in $T_e($OIII), which is often much smaller. However, in the direct determination of $T_e$(OII), its uncertainty are determined by the measured errors of [OII] line fluxes, which can be significantly larger than ones of the [OIII] lines. For example, the estimate of the metallicity for the object Leo P O/H = $(1.5 \pm 0.1) \times 10 ^ {- 5}$ (Aver et al. (2021)) compared to our estimate O/H = $ (1.98 \pm 0.99) \times 10 ^ {- 5}$ (obtained via the direct method) shows $\sim 10$ times less uncertainty, which is a consequence of underestimating the $T_e$(OII) uncertainty. In figure \ref{corr_oh} we present the calculated shifts in the estimates of O/H for the objects from tables \ref{sdss_obj} and \ref{hebcd_obj} obtained using two mentioned methods.

Based on the above, we conclude that (in the context of the determination of Y$_p$) the correct estimate of the HII region metallicity requires $T_e$(OII) and $T_e$(OIII) to be evaluated directly. In turn, the direct method requires using the ratio of the fluxes of the forbidden lines [OII] $ (\lambda 7320 + \lambda 7330) / (\lambda 3726 + \lambda 3729) $ and [OIII] $ \lambda4363 / (\lambda4959 + \lambda 5007)$. This leads to an additional new selection criterion for objects to be analysed - the mentioned oxygen lines must be reliably detected in the spectra of the objects. To determine the physical properties of the interstellar medium (electron temperature and density) and its metallicity, it is preferable to model the statistical and ionization equilibrium in the interstellar medium rather than use semiempirical relations for $T_e$, $n_e$ and O/H.

\noindent
\section{Formation of the sample for the determination of Y$_p $}

\noindent

In order to determine the abundance of the primordial helium considering all the changes mentioned above, we revise the spectroscopic databases described in Kurichin et al. (2021). The SDSS objects database includes 588 objects. From the database we select objects in spectra of which all required lines of helium, hydrogen, and metals are presented. Since the SDSS spectrograph has the wavelength coverage of 3800--9200 \AA \, (Aguado et al. (2019)), only objects with redshifts of 0.020 $ \leq z \leq 0.255 $ are selected for analysis. For such objects the lines [OII] $\lambda3727$ and $\lambda7320 + 7330$ can be detected simultaneously in SDSS spectra. This criterion leaves 161 objects out of 588. Further, from these 161 objects we select ones in the spectra of which the required lines can be reliably detected (a line flux is detected at the $ \geq3 \sigma $ level). It leaves 85 objects for further analysis.

We use the photoionization model (described in detail, Kurichin et al., 2021) to estimate the physical properties and the current abundance of $^4$He (Y) for the selected HII regions. We estimate the temperatures of both ionization zones via the direct method. We determine metallicity of the objects using the PyNeb package (Luridiana et al., 2015). For regression analysis we select objects which are well described by the photoionization model based on the $\chi^2$-criterion ($\chi^2 \leq 4$, which corresponds to a confidence level of 95\% for one degree of freedom). Totally, we select 43 objects for the regression analysis (for a comparison, in our previous work, without using the new criterion, the final sample included 100 objects). Metallicity O/H, the current relative density fraction of $^4$He ($y$ = $n_{\rm He} / n_{\rm H}$), the current $^4$He abundance (Y = $4y / (1 + 4y)\times(1 - Z$)) and $\chi^2$ for the objects are presented in table \ref{sdss_obj}.

In addition to the sample of SDSS objects, we use the HeBCD + NIR spectroscopic database (Izotov et al. 2007, 2014). It contains the optical spectra of 83 HII regions (the HeBCD database), and for some objects there are measured fluxes of the HeI $\lambda10830$ and P$\gamma$ IR lines (Izotov et al. 2014). We also add the extremely metal deficient object Leo P to the sample (the spectrum is taken from Skillman et al. (2013), Aver et al. (2021)). We apply the same selection criteria to the database and get 48 objects in which the physical properties, metallicities, and current abundances of $^4$He are determined. According to the $\chi^2$-criterion, we select 26 objects for further regression analysis, which are presented in table \ref{hebcd_obj}. Note that for objects that have the measured IR line HeI $\lambda10830$, the model has two degrees of freedom. Therefore, for such objects, we use the $\chi^2 \leq 6$ criterion. Totally, the final sample consists of 69 objects (43 objects from the SDSS catalog, 25 objects from the HeBCD + NIR database, plus the Leo P object).

\begin{table*}
\centering
\caption{\label{sdss_obj} Objects from the SDSS database (Kurichin et al. (2021)) selected for the final analysis.}
\scriptsize
\vspace{5mm}
\begin{tabular}{|c|l|c|c|c|c|} 
\hline
№ & Object & O/H $\times 10^5$ & $y$ & Y & $\chi^2$ \\ 
\hline 
1 & J0147+1356 & 7.31 $\pm$ 0.30 & 0.0872 $\pm$ 0.0061 & 0.2582 $\pm$ 0.0133 & 0.70 \\
2 & J0729+3950 & 14.05 $\pm$ 1.43 & 0.0924 $\pm$ 0.0067 & 0.2690 $\pm$ 0.0143 & 0.47 \\
3 & J0806+1949 & 16.85 $\pm$ 1.35 & 0.0875 $\pm$ 0.0034 & 0.2584 $\pm$ 0.0075 & 0.90 \\
4 & J0817+5202 & 23.67 $\pm$ 2.82 & 0.0870 $\pm$ 0.0075 & 0.2569 $\pm$ 0.0165 & 2.93 \\
5 & J0825+3607 & 15.89 $\pm$ 1.28 & 0.0844 $\pm$ 0.0051 & 0.2515 $\pm$ 0.0113 & 2.84 \\
6 & J0840+4707 & 7.60 $\pm$ 0.35 & 0.0862 $\pm$ 0.0040 & 0.2560 $\pm$ 0.0088 & 2.00 \\ 
7 & J0844+0226 & 17.27 $\pm$ 1.40 & 0.0857 $\pm$ 0.0035 & 0.2544 $\pm$ 0.0077 & 2.72 \\
8 & J0845+5308 & 18.63 $\pm$ 1.54 & 0.0860 $\pm$ 0.0070 & 0.2549 $\pm$ 0.0154 & 1.10 \\
9 & J0851+5841 & 6.26 $\pm$ 0.35 & 0.0822 $\pm$ 0.0060 & 0.2471 $\pm$ 0.0135 & 3.32 \\ 
10 & J0907+5327 & 19.76 $\pm$ 2.14 & 0.0815 $\pm$ 0.0046 & 0.2448 $\pm$ 0.0105 & 2.92 \\
11 & J0928+3808 & 13.12 $\pm$ 1.05 & 0.0835 $\pm$ 0.0044 & 0.2498 $\pm$ 0.0098 & 1.19 \\
12 & J0950+0042 & 15.50 $\pm$ 2.21 & 0.0836 $\pm$ 0.0056 & 0.2499 $\pm$ 0.0125 & 2.79 \\
13 & J1024+0525 & 7.73 $\pm$ 0.52 & 0.0829 $\pm$ 0.0036 & 0.2487 $\pm$ 0.0080 & 2.19 \\
14 & J1033+0708 & 29.18 $\pm$ 4.95 & 0.0906 $\pm$ 0.0043 & 0.2644 $\pm$ 0.0092 & 3.93 \\
15 & J1051+1538 & 16.93 $\pm$ 0.69 & 0.0855 $\pm$ 0.0059 & 0.2540 $\pm$ 0.0130 & 3.13 \\ 
16 & J1053+1247 & 14.27 $\pm$ 0.98 & 0.0888 $\pm$ 0.0050 & 0.2614 $\pm$ 0.0108 & 0.89 \\
17 & J1100+4301 & 15.99 $\pm$ 1.19 & 0.0840 $\pm$ 0.0062 & 0.2507 $\pm$ 0.0138 & 3.49 \\
18 & J1105+4445 & 16.28 $\pm$ 1.44 & 0.0844 $\pm$ 0.0039 & 0.2516 $\pm$ 0.0087 & 1.51 \\
19 & J1135+4400 & 17.57 $\pm$ 1.20 & 0.0849 $\pm$ 0.0080 & 0.2526 $\pm$ 0.0178 & 3.11 \\
20 & J1140-0025 & 16.73 $\pm$ 3.04 & 0.0833 $\pm$ 0.0084 & 0.2491 $\pm$ 0.0188 & 1.54 \\
21 & J1143+5330 & 22.67 $\pm$ 2.03 & 0.0892 $\pm$ 0.0073 & 0.2618 $\pm$ 0.0158 & 2.97 \\
22 & J1143+6807 & 18.38 $\pm$ 1.86 & 0.0844 $\pm$ 0.0041 & 0.2515 $\pm$ 0.0091 & 0.29 \\
23 & J1149+3502 & 26.67 $\pm$ 2.83 & 0.0858 $\pm$ 0.0046 & 0.2542 $\pm$ 0.0101 & 3.02 \\
24 & J1200+1343 & 13.78 $\pm$ 0.96 & 0.0894 $\pm$ 0.0046 & 0.2627 $\pm$ 0.0100 & 3.32 \\
25 & J1225+3725 & 6.43 $\pm$ 0.40 & 0.0818 $\pm$ 0.0054 &  0.2463 $\pm$ 0.0122 & 3.83 \\
26 & J1227+5139 & 20.82 $\pm$ 2.55 & 0.0881 $\pm$ 0.0050 & 0.2594 $\pm$ 0.0110 & 3.23 \\
27 & J1249+4743 & 13.23 $\pm$ 0.88 & 0.0823 $\pm$ 0.0062 & 0.2469 $\pm$ 0.0141 & 0.14 \\
28 & J1250+0606 & 18.17 $\pm$ 3.45 & 0.0856 $\pm$ 0.0082 & 0.2542 $\pm$ 0.0182 & 2.17 \\
29 & J1301+1240 & 20.04 $\pm$ 2.63 & 0.0893 $\pm$ 0.0050 & 0.2621 $\pm$ 0.0109 & 1.24 \\
30 & J1322+0130 & 22.26 $\pm$ 3.12 & 0.0869 $\pm$ 0.0048 & 0.2568 $\pm$ 0.0105 & 2.59 \\
31 & J1335+0414 & 21.04 $\pm$ 2.38 & 0.0853 $\pm$ 0.0058 & 0.2534 $\pm$ 0.0128 & 2.86 \\
32 & J1347+6202 & 15.04 $\pm$ 2.78 & 0.0818 $\pm$ 0.0077 & 0.2458 $\pm$ 0.0175 & 1.58 \\ 
33 & J1424+2257 & 5.59 $\pm$ 0.59 & 0.0827 $\pm$ 0.0063 & 0.2483 $\pm$ 0.0142 & 2.78 \\
34 & J1426+6300 & 18.54 $\pm$ 1.77 & 0.0839 $\pm$ 0.0061 & 0.2504 $\pm$ 0.0136 & 1.80 \\
35 & J1430+0643 & 21.77 $\pm$ 2.03 & 0.0861 $\pm$ 0.0047 & 0.2551 $\pm$ 0.0103 & 2.16 \\
36 & J1448-0111 & 12.82 $\pm$ 0.93 & 0.0889 $\pm$ 0.0055 & 0.2617 $\pm$ 0.0120 & 2.50 \\
37 & J1509+3732 & 6.48 $\pm$ 0.30 & 0.0827 $\pm$ 0.0031 & 0.2483 $\pm$ 0.0070 & 3.66 \\
38 & J1510+3732 & 6.50 $\pm$ 0.32 & 0.0845 $\pm$ 0.0035 & 0.2522 $\pm$ 0.0078 & 2.81 \\
39 & J1517-0008 & 25.59 $\pm$ 3.65 & 0.0846 $\pm$ 0.0072 & 0.2515 $\pm$ 0.0160 & 0.19 \\
40 & J1538+5842 & 22.44 $\pm$ 2.57 & 0.0848 $\pm$ 0.0063 & 0.2522 $\pm$ 0.0140 & 2.87 \\
41 & J1557+2321 & 20.50 $\pm$ 1.88 & 0.0865 $\pm$ 0.0034 & 0.2561 $\pm$ 0.0075 & 2.95 \\
42 & J2115-0800 & 19.21 $\pm$ 1.79 & 0.0889 $\pm$ 0.0061 & 0.2614 $\pm$ 0.0133 & 2.26 \\
43 & J2329-0110 & 17.96 $\pm$ 2.45 & 0.0856 $\pm$ 0.0097 & 0.2542 $\pm$ 0.0215 & 3.51 \\
\hline
\end{tabular}
\end{table*}

\begin{table*}
\centering
\scriptsize
\caption{\label{hebcd_obj} Objects from the HeBCD + NIR database (Izotov et al., 2007, 2014) and Leo P (Aver et al., 2021) selected for the final analysis.}
\vspace{5mm}
\begin{tabular}{|c|l|c|c|c|c|} 
\hline
№ & Object & O/H $\times 10^5$ & $y$ & Y & $\chi^2$\\ 
\hline 
1 & HS 0122+0743 & 4.32 $\pm$ 0.28 & 0.0874 $\pm$ 0.0044 & 0.2589 $\pm$ 0.0097 & 0.69 \\
2 & HS 0811+4913 & 9.66 $\pm$ 0.56 & 0.0809 $\pm$ 0.0036 & 0.2441 $\pm$ 0.0081 & 1.79 \\
3 & HS 1213+3636A & 23.81 $\pm$ 2.97 & 0.0899 $\pm$ 0.0032 & 0.2632 $\pm$ 0.0068 & 3.38 \\
4 & HS 1214+3801 & 10.76 $\pm$ 0.48 & 0.0907 $\pm$ 0.0040 & 0.2656 $\pm$ 0.0086 & 2.31 \\
5 & HS 2359+1659 & 14.93 $\pm$ 0.85 & 0.0838 $\pm$ 0.0055 & 0.2502 $\pm$ 0.0123 & 2.35 \\
6 & I Zw 18 SE\footnotemark[1] & 1.31 $\pm$ 0.14 & 0.0779 $\pm$ 0.0032 & 0.2375 $\pm$ 0.0074 & 0.36 \\
7 & Leo P\footnotemark[2] & 1.98 $\pm$ 0.77 & 0.0827 $\pm$ 0.0043 & 0.2484 $\pm$ 0.0097 & 1.61 \\
8 & Mrk 59\footnotemark[1] & 9.71 $\pm$ 0.37 & 0.0857 $\pm$ 0.0031 & 0.2547 $\pm$ 0.0069 & 0.80 \\
9 & Mrk 71\footnotemark[1] & 6.77 $\pm$ 0.22 & 0.0862 $\pm$ 0.0028 & 0.2560 $\pm$ 0.0061 & 1.68 \\
10 & Mrk 209\footnotemark[1] & 5.88 $\pm$ 0.22 & 0.0820 $\pm$ 0.0019 & 0.2467 $\pm$ 0.0044 & 0.23 \\
11 & Mrk 450\footnotemark[1] & 13.90 $\pm$ 0.64 & 0.0860 $\pm$ 0.0032 & 0.2553 $\pm$ 0.0070 & 3.01 \\
12 & Mrk 1315\footnotemark[1] & 18.88 $\pm$ 0.77 & 0.0876 $\pm$ 0.0016 & 0.2586 $\pm$ 0.0034 & 1.13 \\
13 & Mrk 1329\footnotemark[1] & 20.27 $\pm$ 1.23 & 0.0892 $\pm$ 0.0035 & 0.2620 $\pm$ 0.0076 & 3.78 \\
14 & SBS 0335--052 E\footnotemark[1] & 2.10 $\pm$ 0.10 & 0.0848 $\pm$ 0.0027 & 0.2532 $\pm$ 0.0061 & 1.03 \\
15 & SBS 0917+527 & 7.97 $\pm$ 0.76 & 0.0817 $\pm$ 0.0033 & 0.2459 $\pm$ 0.0075 & 0.51 \\
16 & SBS 0940+544 2\footnotemark[1] & 3.31 $\pm$ 0.23 & 0.0850 $\pm$ 0.0031 & 0.2535 $\pm$ 0.0068 & 0.37 \\
17 & SBS 0946+558 & 11.46 $\pm$ 0.83 & 0.0834 $\pm$ 0.0030 & 0.2495 $\pm$ 0.0067 & 1.61 \\
18 & SBS 1030+583\footnotemark[1] & 5.99 $\pm$ 0.41 & 0.0813 $\pm$ 0.0025 & 0.2450 $\pm$ 0.0056 & 1.29 \\
19 & SBS 1054+365 & 8.92 $\pm$ 0.66 & 0.0878 $\pm$ 0.0034 & 0.2594 $\pm$ 0.0075 & 0.69 \\
20 & SBS 1135+581\footnotemark[1] & 10.87 $\pm$ 0.40 & 0.0851 $\pm$ 0.0011 & 0.2534 $\pm$ 0.0024 & 4.70 \\
21 & SBS 1152+579\footnotemark[1] & 7.00 $\pm$ 0.25 & 0.0824 $\pm$ 0.0046 & 0.2476 $\pm$ 0.0103 & 1.49 \\
22 & SBS 1211+540 & 4.73 $\pm$ 0.39 & 0.0804 $\pm$ 0.0027 & 0.2431 $\pm$ 0.0061 & 3.97 \\
23 & SBS 1222+614\footnotemark[1] & 8.08 $\pm$ 0.36 & 0.0849 $\pm$ 0.0027 & 0.2531 $\pm$ 0.0059 & 3.27 \\
24 & Tol 1214--277\footnotemark[1] & 3.72 $\pm$ 0.27 & 0.0836 $\pm$ 0.0023 & 0.2504 $\pm$ 0.0052 & 2.48 \\
25 & Tol 65\footnotemark[1] & 3.21 $\pm$ 0.15 & 0.0811 $\pm$ 0.0029 & 0.2448 $\pm$ 0.0065 & 0.60 \\
26 & UM 311\footnotemark[1] & 18.13 $\pm$ 2.53 & 0.0847 $\pm$ 0.0018 & 0.2522 $\pm$ 0.0041 & 0.29 \\
\hline
\multicolumn{5}{l}{{{}}}\\
\multicolumn{5}{l}{\footnotemark[1] {\scriptsize{This object has a measured flux of IR line $\lambda 10830$ presented in Izotov et al. (2014).}}}\\
\multicolumn{5}{l}{\footnotemark[2] {\scriptsize{The spectrum of this object is taken from Aver et al. (2021).}}}
\end{tabular}
\end{table*}

\section*{Regression analysis and results}

\begin{figure}
\begin{center}
        \includegraphics[width = 1.1\columnwidth]{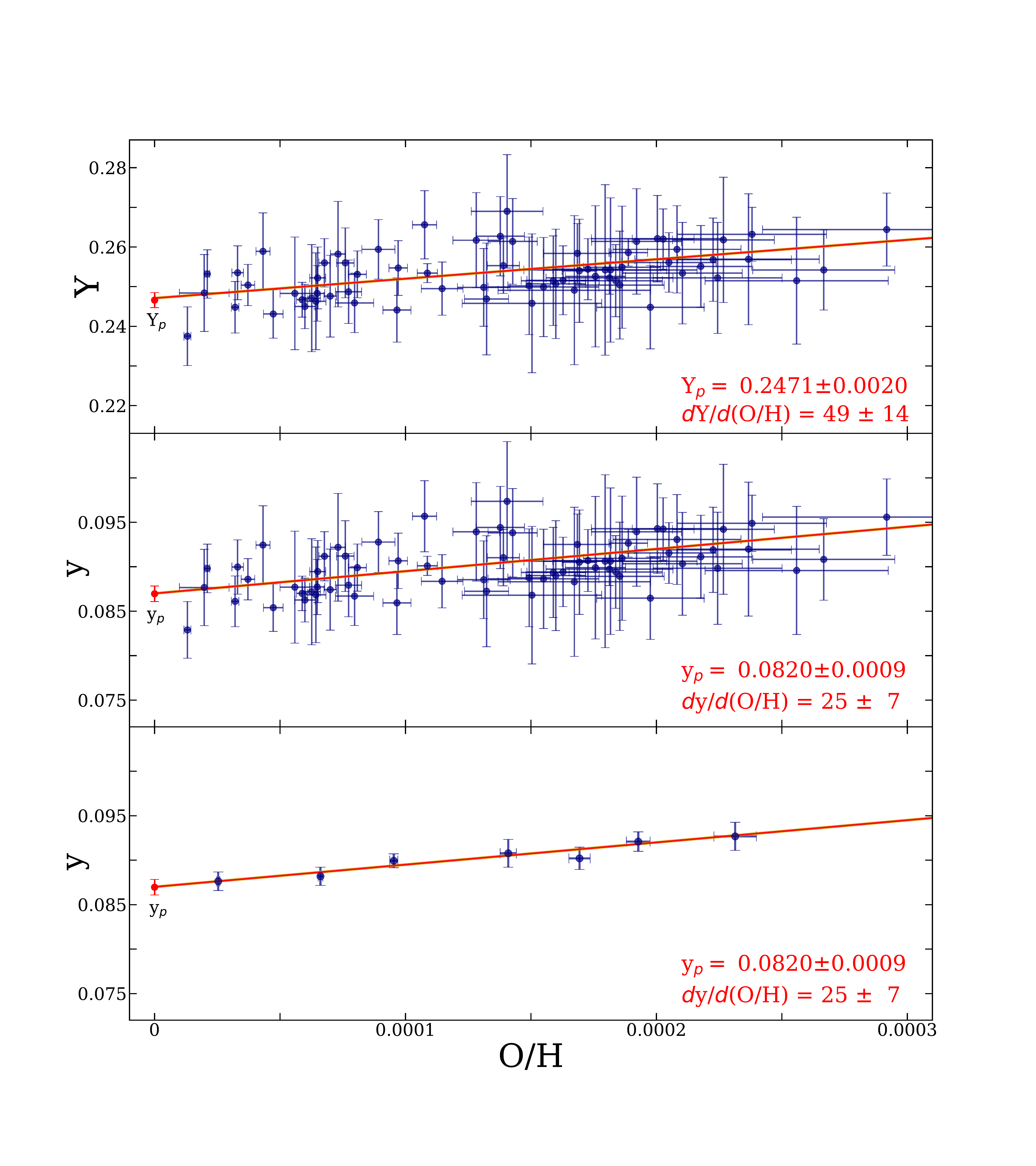}
        \caption{The Y--O/H (the upper panel) and $ y $--O/H (the center panel) diagrams for the final sample (tables \ref{sdss_obj} and \ref{hebcd_obj}). The lower panel shows the same regression line as the central panel, but the entire sample is divided into equally sized bins, in each of those the weighted mean point is calculated for a better visualisation.}
        \label{regression}
\end{center}
\end{figure}

The abundance of $^4$He for 69 objects from the final sample is shown in Fig. \ref{regression} along with the estimated metallicities of these objects (the upper panel). For these objects we preform Y - O/H regression analysis using the following relation:
\begin{equation}
    {\rm Y} = \rm{Y}_p + \frac{d{\rm Y}}{d({\rm O/H})} \times {\rm O/H}
\end{equation}

We use the Markov Chain Monte Carlo method (MCMC) to estimate values and uncertainties in the regression parameters. We obtain the following estimates for Y$_p$ and the slope $d$Y/$d$(O/H):
\begin{equation}
    {\rm Y}_p = 0.2471 \pm 0.0020 ~~~~\text{and}~~~~ \frac{d{\rm Y}}{d{\rm (O/H)}} = 49 \pm 14
\end{equation}

In the recent paper Hsyu et al. (2020), the authors proposed to use the ratio of the helium-to-hydrogen volume density $y$ instead of the mass abundance Y. These quantities are related via the following formula:
\begin{equation}
    {\rm Y} = \frac{4 y}{1 + 4 y} \times (1 - Z)
\end{equation}
where $ Z $ is the total metallicity of an object. The relation between O/H and $Z$ is given by the formula $Z = $ C $\times$ O/H. In works devoted to the determination of Y$_p$, C = 20 is usually taken for all studied objects (Pagel et al. (1992)). The coefficient C is determined by the rate of the chemical evolution of each particular galaxy, and therefore, in fact, it should be different for each object. The use of $y$ instead of Y allows one to eliminate the described dependence on the unknown value of $Z$. Then, in the regression analysis for $y$ - O/H, the $y_p$ value will be estimated, which does not include a model dependence on $Z$. The quantity $y_p$ is related to Y$_p$ via the following formula:
\begin{equation}
    {\rm Y}_p = \frac{4 {\rm y}_p }{1 + 4 {\rm y}_p}
    \label{y-Y}
\end{equation}

The central panel of figure \ref{regression} shows the $y$ - O/H diagram for the SDSS and HeBCD objects. Preforming the regression analysis of the sample, we obtain the following results:
\begin{equation}
    {\rm y}_p = 0.0820 \pm 0.0009 ~~~~\text{and}~~~~ \frac{d{\rm y}}{d{\rm (O/H)}} = 25 \pm 7
\end{equation}

After converting y$_p$ to Y$_p$ using the formula \ref{y-Y}, we get the following estimate of Y$_p$:
\begin{equation}
    {\rm Y}_p = 0.2470 \pm 0.0020
\end{equation}

The results are in good consistency with our previous estimate Y$_p = 0.2462 \pm 0.0022$. Despite the fact that the data sample used in this paper - 69 objects - is smaller than 120 objects used in our previous work (Kurichin et al. (2021)), we  have estimated the regression parameters with better precision. This is because the more correct determination of O/H for the objects allowed to significantly reduce the intrinsic scatter in the sample, which leads to more tight parameter constraints. The obtained estimate is also in good agreement with the estimates of Y$_p$ obtained in other analogous works (see table \ref{tab:helium_dets}).

\section*{Conclusion}
\noindent

In this paper we have analysed the methods of the determination of HII region metallicity, which are currently used in various works concerned with the determination of the primordial $^4$He abundance.
\begin{itemize}
\item We have shown that the use of different lines of the oxygen ion OII for the determination of the OII/H abundance gives inconsistent results for all investigated methods. The reason for this discrepancy is the incorrect determination of the temperature of the low ionization zone in HII regions ($T_e$(OII)). In the most of the works this quantity have been estimated indirectly using the temperature of the high ionization zone $T_e$(OIII).

\item We have shown that the use of the relations $T_e$(OII) = $f(T_e$(OIII)) biases the estimate of the metallicity and leads to an incorrect determination of its uncertainty.

\item
We have shown that to obtain a correct estimate of OII/H, $T_e$(OII) should be evaluated via the direct method (similar to the one for the determination of $T_e$(OIII)) using the line ratio [OII] $(\lambda3726 + \lambda3729) / (\lambda7320 + \lambda7330)$. This requirement implies a new additional selection criterion for the objects to be used in the determination of Y$_p$: the mentioned [OII] lines have to be confidently detected in the spectra of the studied objects. 
\end{itemize}

Applying the new selection criterion along with the criteria described in our previous paper (Kurichin et al., 2021) to the SDSS database (Kurichin et al., 2021) and the HeBCD+NIR database (Izotov et al., 2007, 2014), we have selected 69 objects for the determination of Y$_p$. We have preformed the regression analysis of the final sample and obtained the estimate Y$_p = 0.2470 \pm 0.0020$, which is in good consistency with our previous estimate Y$_p = 0.2462 \pm 0.0022$ (Kurichin et al., 2021), as well as with other independent estimates (see table 1). In the present paper we  have used almost twice smaller sample of objects for the regression analysis compared to our previous paper. However, it is important to note that the accuracy of estimate of Y$_p$, obtained in the present paper, is higher compared to one from Kurichin et al. (2021). This is due to the fact that the correct determination of the metallicity (via the direct method) lowers the intrinsic scatter in the regression sample. In turn, it allows to put more tight constraints on the estimates of Y$_p$ and $d$Y/$d$(O/H). 

Our estimate is also in a good agreement with the estimate obtained using the numerical codes of Primordial Nucleosynthesis in combine with the analysis of CMB anisotropy: Y$_p = 0.2470 \pm 0.0002$ (Planck Collaboration, 2020). It is important for checking the self-consistency of the Standard Cosmological Model, since the two obtained estimates refer to different cosmological eras (a possible discrepancy between them could indicate a new physics).

Further analysis of data from the SDSS catalog using the proposed method will sufficiently enhance the estimate of Y$_p$. Potentially, it may allow to achieve the accuracy of the estimate of $\eta = n_b/n_\gamma$ comparable with the accuracy obtained in works concerned with the determination of the primordial deuterium abundance.

\section*{Aknowlegments}

The work is supported by the Russian Science Foundation (grant 18-12-00301).

\section*{References}

E. Aver, K.A. Olive and E.D. Skillman, J. Cosmol. Astropart. Phys 07, 011 (2015). \\
E. Aver, D.A. Berg, K.A. Olive et, al, J. Cosmol. Astropart. Phys.  03, 027 (2021). \\
D. S. Aguado, Romina Ahumada, Andrés Almeida, et al, Astron. Astrophys. Suppl. Ser. 240, 23 (2019).\\ 
S.A. Balashev, E.O. Zavarygin, A.V. Ivanchik, et al, Mon. Not. R. Astron. Soc. 458, 2188 (2016). \\
M. Valerdi, A. Peimbert, eprint arXiv:1905.05102 (2019). \\
M. Valerdi, A. Peimbert, M. Peimbert, Mon. Not. R. Astron. Soc. 505, 3624 (2021). \\
E.O. Zavarygin, J.K. Webb, V. Dumont, et al., Mon. Not. R. Astron. Soc. 477, 5536 (2018).\\ 
Y.I. Izotov,T.X. Thuan, V.A. Lipovetsky, Astrophys. J. 435, 647 (1994).\\
Y.I. Izotov, G. Stasińska, G. Meynet, et al., Astron. Astrophys. 448, 955 (2006).\\
Y.I. Izotov, T.X. Thuan, G. Stasińska, Astrophys. J. 662, 15 (2007).\\
Y.I. Izotov, T.X. Thuan, N.G. Guseva, Mon. Not. R. Astron. Soc. 445, 778 (2014).\\
Planck Collaboration, et al., Astron. Astrophys. 641, A6 (2020).\\
R.J. Cooke, M. Pettini, R.A. Jorgenson, et al., Astrophys. J. 781, 31 (2014). \\
R.J. Cooke, M. Pettini, C.C. Steidel, Astrophys. J. 855, 102 (2018). \\
R.J. Cooke, M. Fumagalli, Nature Astronomy 2, 957 (2018). \\
O.A Kurichin, P.A. Kislitsyn, V.V. Klimenko, et al., Mon. Not. R. Astron. Soc. 502, 3045 (2021). \\
Luridiana, V., Morisset, C., Shaw, R. A., Astron. Astrophys. 573, A42 (2015). \\
P. Noterdaeme, S. López, V. Dumont, et al., Astron. Astrophys. 542, L33 (2012). \\
L.S. Pilyugin, J.M. V´ılchez, B. Cedres et al., Mon. Not. R. Astron. Soc. 403 896-905 (2010).\\
B.E.J. Pagel, E.A. Simonson, R.J. Terlevich et al.,
Mon. Not. R. Astron. Soc. 255, 325 (1992).\\
E. Pérez-Montero, Publ. Astron. Soc. Pac. 129, 974 (2017).\\
E. Pérez-Montero, A.I. Díaz, Mon. Not. R. Astron. Soc. 346, 105 (2003).\\
A. Peimbert, M. Peimbert, V. Luridiana,
Rev. Mex. Astron. Astrofís. 52, 419 (2016).\\
S. Riemer-Sørensen, S. Kotuš, J.K. Webb, et al., Mon. Not. R. Astron. Soc. 468, 3239 (2017).\\
E.D. Skillman, J.J. Salzer, D.A. Berg, et al., Astron. J. 146, 3 (2013).\\
T. Hsyu, R.J. Cooke, J.X. Prochaska, et al., Astrophys. J. 896, 77 (2020).\\
V. Fernández, E. Terlevich, A.I. Díaz, et al.,  Mon. Not. R. Astron. Soc. 478, 5301 (2018).\\
V. Fernández, E. Terlevich, A.I. Díaz, et al., Mon. Not. R. Astron. Soc. 487, 3221 (2019).\\
G.F. Hägele, E. Pérez-Montero, A.I. Díaz, et al., Mon. Not. R. Astron. Soc. 372, 293 (2006).\\
G.F. Hägele, A.I. Díaz, E. Terlevich, et al., Mon. Not. R. Astron. Soc. 383, 209 (2007).\\
Tsivilev, A. P., Parfenov, S. Yu., Sobolev, A. M., et al., Astron. Lett. 39, 11 (2013).\\

\end{document}